\documentclass[12pt]{iopart}

\usepackage[dvips]{graphicx}

\begin{document}

\title{Performance of the WaveBurst algorithm on LIGO data}

\author{S~Klimenko\dag, I~Yakushin\ddag, M~Rakhmanov\dag and
G~Mitselmakher\dag} 

\address{\dag\ Department of Physics, University of Florida, Gainesville,
FL~32611, USA} 
\address{\ddag\ LIGO Livingston Observatory, Livingston, LA~70754, USA} 

\begin{abstract}
In this paper we describe the performance of the WaveBurst algorithm
which was designed for detection of gravitational wave bursts in
interferometric data. 
The performance of the algorithm was evaluated on the test data set collected
during the second LIGO Scientific run. We have measured the false alarm rate 
of the algorithm as a function of the threshold and estimated  
its detection efficiency for simulated burst waveforms.
\end{abstract}

\maketitle

\section{Introduction}

A direct observation of gravitational waves (GW) produced by astrophysical
sources is an ultimate goal for a new generation of detectors based on 
laser interferometry \cite{LIGO}-\cite{TAMA}.
A broad class of expected signals are bursts which are transients 
of gravitational radiation with short duration, typically less then a second. 
They may be produced by sources such as supernova 
explosions, mergers of binary inspiral systems, gamma ray
bursts, and other violent and energetic phenomena in the universe.
The first result of burst searches with LIGO detectors has been 
recently reported by the LIGO collaboration \cite{LIGO_burst}.

At the present time the waveforms of gravitational waves from burst
sources are poorly 
known. Consequently, burst searches employ data analysis 
algorithms which identify bursts with a broad range 
of possible waveforms. Recently several such algorithms have been developed 
 \cite{tf1,tf2,power,tf3}, including the algorithm called 
WaveBurst \cite{method}.
The WaveBurst algorithm is based on wavelet transformations 
and allows detection of a wide class of 
GW bursts by using a bank of wavelet packets \cite{vidakovic,packets1}.

In this paper we describe a data analysis pipeline with 
a particular implementation of the WaveBurst algorithm and evaluate 
the pipeline performance on the data from
the second LIGO Scientific run (S2).
The S2 data was collected from all three LIGO detectors
during a two month period beginning on February 14, 2003.
We used a small fraction ($\sim{10}\%$) of the S2 data, which was selected 
as the test (playground) set for the purpose of tuning and evaluation
of the burst detection algorithms.
The paper is organized as follows: we begin with a brief description
of the data analysis pipeline, present
the measurements of the false alarm rates, and conclude with the estimation
of the pipeline sensitivity.

\section{WaveBurst data analysis pipeline}

\begin{figure}[ht]
  \begin{center}
  \includegraphics[width=0.7\textwidth]{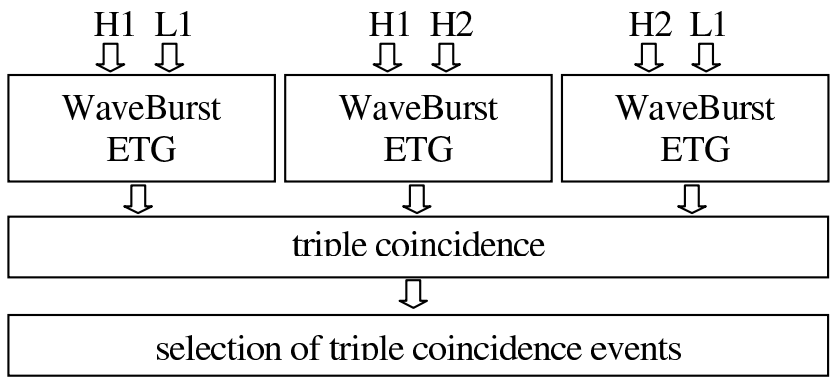}
  \end{center}
  \caption{The block diagram of the WaveBurst data analysis pipeline for
  three LIGO detectors: H1, H2 and L1.}
  \label{fig:pipeline}
\end{figure}

The block diagram of the WaveBurst data analysis pipeline is shown in 
Figure~\ref{fig:pipeline}. 
The pipeline takes as input the raw time series from the GW channels of the
three LIGO detectors  and generates a set of potential GW burst events (triggers). 
The data processing stages of the pipeline are:
\begin{itemize}
   \item{production of burst triggers with the WaveBurst event
   trigger generator (ETG),}
   \item{reconstruction of events coincident in all three LIGO detectors,} 
   \item{final selection of the triple coincidence events.}
\end{itemize}

\subsection{Production of burst triggers}
\label{s_ETG}

The WaveBurst ETG is an implementation of the burst analysis method
described in \cite{method}. The method is based on wavelet
transformations which allow time-frequency  
representation of data. The bursts are identified by looking for regions 
in the wavelet domain with an excess of power inconsistent with stationary 
detector noise. 

The ETG takes data from the GW channels of two detectors and
produces a list of coincident burst triggers. The data is processed 
in the following steps:
1) wavelet transformation, 
2) selection of wavelet amplitudes,
3) coincidence between the channels,
4) generation of burst triggers, and
5) selection of burst triggers. 
During steps 1), 2) and 4) the data processing is independent for
each channel. 
During  steps 3) and 5) data from both channels is used.

The input data are time series with duration of 120 seconds and
sampling rate of $8192$~Hz.  Using the orthogonal wavelet   
transformation\footnote{SYMLET wavelet with the filter length of 60.}
the time series are converted into wavelet series $W_{ij}$, where
$i$ is the time index and $j$ is the wavelet layer index. Each wavelet
layer can be associated with a certain frequency band of the initial time 
series. Therefore, the wavelet series $W_{ij}$ can be displayed as a time-frequency
scalogram consisting of 64 wavelet layers with $n=15360$ pixels (data samples) 
each. The time-frequency resolution of the WaveBurst scalograms is 
the same for all wavelet layers ($1/128$~sec $\times \; 64$~Hz), which is 
different from dyadic time-frequency resolution of conventional wavelets 
\cite{method,ingrid}. The constant time-frequency resolution makes the 
WaveBurst scalograms similar to spectrograms produced with windowed Fourier 
transformations.

For each layer
we select a fixed fraction ($P$) of pixels with largest absolute 
amplitudes, which are called {\it black pixels}. The number of black pixels is
$nP$. All other wavelet pixels are called {\it white pixels}. 
Then we calculate rank statistics for the black pixels. The rank $R_{ij}$ is an 
integer number from 1 to $nP$ with the rank 1 assigned to the pixel with the largest 
amplitude. Given the rank $R_{ij}$, the following non-parametric
statistic is computed 
\begin{equation}\label{eq:rsig}
   y_{ij} = -\ln \left( \frac{R_{ij}}{nP} \right).
\end{equation}
For white pixels the value of $y_{ij}$ is set to zero.
The statistic $y_{ij}$ has a meaning of the pixel {\it logarithmic significance}. 
Assuming Gaussian detector noise, the logarithmic significance can be also calculated as
\begin{equation}\label{eq:pL}
   \tilde{y}_{ij} = g_{P}(\tilde{w}_{ij}) = \ln(P) - \ln \left( 
     \sqrt{2/\pi} \int_{\tilde{w}_{ij}}^{\infty}  
     \e^{-x^2/2} \; dx \right) ,
\end{equation}
where $\tilde{w}_{ij}$ is the absolute value of the pixel amplitude in units of
the noise standard deviation. In practice, the LIGO detector noise is not Gaussian and 
its probability  distribution function is not well known. Therefore, we use the
non-parametric statistic $y_{ij}$, which is a more robust measure of the pixel significance
than $\tilde{y}_{ij}$. 
Using the inverse function of $g_P$ with $y_{ij}$ as an argument,
we introduce the {\it non-parametric amplitude}
\begin{equation}\label{eq:amp}
   w_{ij} = g_P^{-1}(y_{ij}),
\end{equation}
and the {\it excess power ratio} 
\begin{equation}\label{eq:amp}
   \rho_{ij} = w^2_{ij}-1,
\end{equation}
which characterizes the pixel excess power above the average detector noise.

After the black pixels have been selected, 
we require a coincidence between the channels. Given a black pixel in 
the first channel, it is accepted or rejected depending on 
a condition applied to pixels in the corresponding
time-frequency area of the second channel. The pixel is accepted if the 
significance of pixels in the second channel satisfies
\begin{equation}
   y_{(i-1)j}+y_{ij}+y_{(i+1)j} > \eta,
\end{equation}
where $\eta$ is the {\it coincidence threshold}. Otherwise, the pixel 
is rejected. This  procedure is repeated for all black pixels in the first channel. 
The same coincidence algorithm is applied to pixels in the second channel.
As a result, a considerable number of black pixels in both channels
produced by fluctuations of the detector noise is rejected. 
At the same time, black pixels produced by coincident bursts have a much higher
acceptance probability because of the coherent excess of power in two
detectors.

After the coincidence procedure is applied to both channels, the channels 
may have similar but not identical patterns of black pixels which form
clusters. The clusters can be reconstructed separately for each
channel. However, in the present WaveBurst algorithm, we merge the
black pixels from both channels into one time-frequency scalogram and
then run a cluster analysis. For each black pixel 
we define neighbors, which share a side or a vertex with the black pixel. 
Neighbors can be both black and white pixels. The white 
neighbors are called {\it halo} pixels. We define a cluster as a connected
group of black and halo pixels. After the cluster reconstruction, we go back to the
original time-frequency scalograms and  calculate the cluster parameters
separately for each channel. There are always two
clusters, one per channel, which form a WaveBurst trigger.

The cluster parameters are calculated using black pixels only. 
For example, the cluster size $k$ is defined as the number of black 
pixels. Other parameters, which characterize the cluster 
strength, are the cluster 
excess energy ratio $\rho$ and the cluster {\it logarithmic likelihood} $Y_k$: 
\begin{equation}
   \rho = \sum_{i,j\in{C(k)}} \rho_{ij},~~Y_k = \sum_{i,j\in{C(k)}} y_{ij},
\end{equation}
where $C(k)$ defines a set of black pixels in the cluster.
The cluster size and the excess power ratio are used for 
the selection of triggers. A trigger is reported by the ETG if both clusters 
satisfy the conditions: $k>0$ and $\rho>6.25$. 

\subsection{ETG tuning}
\label{tuning}
There are two main ETG parameters: 
the black pixel fraction $P$ and the coincidence threshold $\eta$.
The purpose of these parameters is to control the average black pixel 
occupancy $O(P,\eta)$ of the scalograms used for the cluster reconstruction.
To ensure robust cluster reconstruction, the occupancy should not be greater than $1$\%.
For white Gaussian detector noise the functional dependence $O(P,\eta)$ can be 
calculated analytically. We require that $O(P,\eta)=0.7\%$, which sets a constraint 
on $P$ and $\eta$. 

The selection of black pixels effectively sets a threshold on the wavelet amplitudes:
$ w_{ij} > g_P^{-1}(0)$. The larger the value of $P$ -- the lower the threshold.
However, the black pixel fraction should not be greater than 31.7\%, otherwise 
pixels with negative values of $\rho_{ij}$ would be taken into the analysis.
From the other side, with $P$ set too small (less than a few percent),
noise outliers due to instrumental glitches may consume 
the entire time-frequency volume
available for black pixels and thus mask gravitational waves. 
To avoid saturation from the instrumental glitches, we run the analysis 
with $P$ equal to 10\%. Together with 
the occupancy constraint above it sets the coincidence threshold of $1.5$.

For S2 playground data and selected values of $P$ and $\eta$, the average 
trigger rate 
is approximately $6$~Hz. Within a factor of two it is consistent with the false alarm 
rate expected for the white Gaussian detector noise. 
To reduce the ETG rates to manageable 
level ($\sim{1}$~Hz), we set an intermediate threshold on the cluster excess power 
ratio (Section~\ref{s_ETG}) and postpone the final selection of the pipeline threshold 
for the post-production analysis (Section~\ref{final}). 
In this case, there is no need to 
re-run the ETG, which is CPU and time consuming operation,  
for tuning of the pipeline false alarm rates and sensitivity.

\subsection{Triple coincidence}

The output of the WaveBurst ETG is a set of double coincidence triggers
for a selected interferometer pair $X\times{Y}$. 
For three LIGO interferometers there are three possible 
pairs: $L1\times{H1}$, $H1\times{H2}$ and 
$H2\times{L1}$, where $H1$ and $H2$  are two Hanford interferometers
and $L1$ is the Livingston interferometer. 
To identify triple coincidence events, we require  
a time-frequency coincidence of the WaveBurst triggers generated for
these three pairs. 

For time coincidence the following conditions are required
\begin{equation}
   |T_{L1H1}-T_{H1H2}|<T_w,~|T_{H2L1}-T_{H1H2}|<T_w,~|T_{H2L1}-T_{L1H1}|<T_w,
\end{equation}
where $T_w$ is the time window ($T_w=20~\mathrm{ms}$) and $T_{XY}$ is
the average center time of the $X$ and $Y$ clusters. 
Given the time $t_i$ for individual pixels, the cluster center time is calculated as 
\begin{equation}
   T = {\sum_{i,j\in{C(k)}} {t_i \; w^2_{ij}}} \; / {\sum_{i,j\in{C(k)}} {w^2_{ij}}}.
\end{equation}

We also apply a loose requirement on the frequency coincidence of
the WaveBurst triggers. 
First, we calculate the minimun ($f_{min}$) and maximum ($f_{max}$)
frequency for each interferometer pair $X\times{Y}$
\begin{equation}
   f_{min} = min(f_{low}^X,f_{low}^Y), \qquad 
   f_{max} = max(f_{up}^X,f_{up}^Y),
\end{equation}
where $f_{low}$ and $f_{up}$ are the lower and upper frequency
boundaries of the $X$ and $Y$ clusters. 
Then the trigger frequency bands are calculated as $f_{max}-f_{min}$
for all pairs. For frequency coincidence, the bands of all three WaveBurst triggers are 
required to overlap.

\subsection{Final selection of triple coincidence events}
\label{final}

\begin{figure}[ht]
  \begin{center}
  \includegraphics[width=0.7\textwidth]{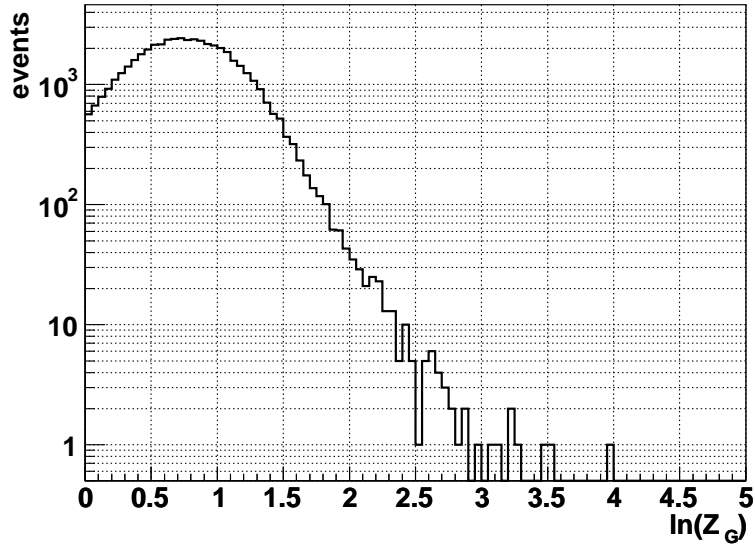}
  \end{center}
  \caption{The combined significance distribution of the WaveBurst
  triggers produced with unphysical time lags between the Livingston
  and Hanford detectors.}
  \label{fig:GCGA}
\end{figure}

The triple coincidence events consist of 3 WaveBurst triggers or 6 clusters. 
The cluster parameters are
calculated separately for each interferometer, as described in Section~\ref{s_ETG}. 
For the final selection of the burst triggers we use the {\it cluster significance}
\begin{equation}
   Z = 
   Y_k - \ln{ \left( \sum_{m=0}^{k-1}\frac{Y_k^{m}}{m!} \right) }
\end{equation}
derived from the logarithic likelihood $Y_k$ \cite{method}. 
Given the significance of six clusters, we compute the
{\it combined significance} of the triple coincidence event as
\begin{equation}
   Z_G = \left( Z^{L1}_{L1H1} \; Z^{H1}_{L1H1} \; Z^{H2}_{H2L1} \; 
                Z^{L1}_{H2L1} \; Z^{H1}_{H1H2}~\; Z^{H2}_{H1H2} 
         \right)^{1/6},
\end{equation}
where $Z^{X}_{XY}$ ($Z^{Y}_{XY}$) is the significance of $X$ ($Y$) cluster for the $X \times Y$
interferometer pair.
Figure~\ref{fig:GCGA} shows the combined significance distribution for the 
false alarm events found in the S2 playground data.
To control the pipeline false alarm rate, we set a threshold on the
value of the combined significance. 

\section{False alarm rate}

\begin{figure}[ht]
  \begin{center}
  \includegraphics[width=0.7\textwidth]{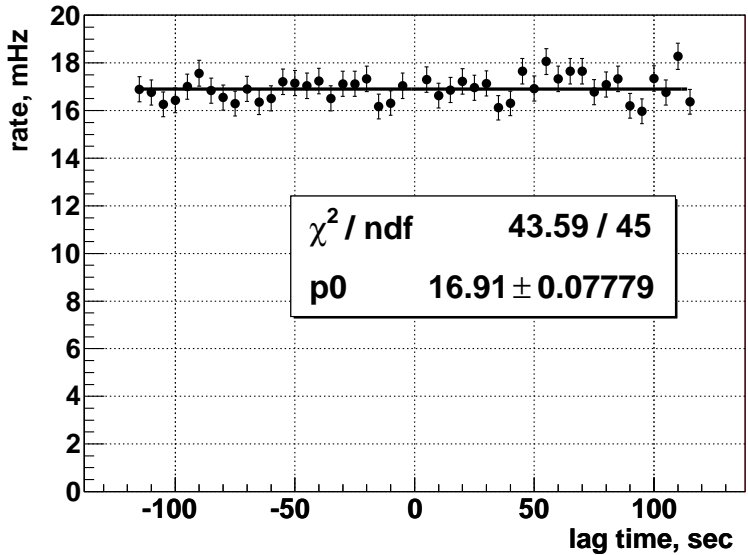}
  \end{center}
  \caption{The false alarm rate before the $Z_G$ selection cut as a function
  of the time lag between the Livingston and Hanford detectors.}
  \label{fig:bg_lags}
\end{figure}

Assuming there is no physical correlation between the Hanford and Livingston sites, 
the false alarm (FA) rate is dominated by accidental triple-coincidence events 
produced by fluctuations of the detector noise. 
For estimation of the FA rate we perform the entire WaveBurst analysis on
the data with  
artificial time shifts between the Livingston and Hanford detectors. 
The false alarm triggers are generated for 46 time lags between $-115$ and 
$115$ seconds with time step of $5$ seconds. The zero time lag is
excluded from the analysis.
Figure~\ref{fig:bg_lags} shows the measured rates at different time lags 
before the final selection cut is applied to the combined significance.
The measurements at different time lags are consistent with the average rate
of $16.9$ Hz.

\begin{figure}[ht]
  \begin{center}
  \includegraphics[width=0.7\textwidth]{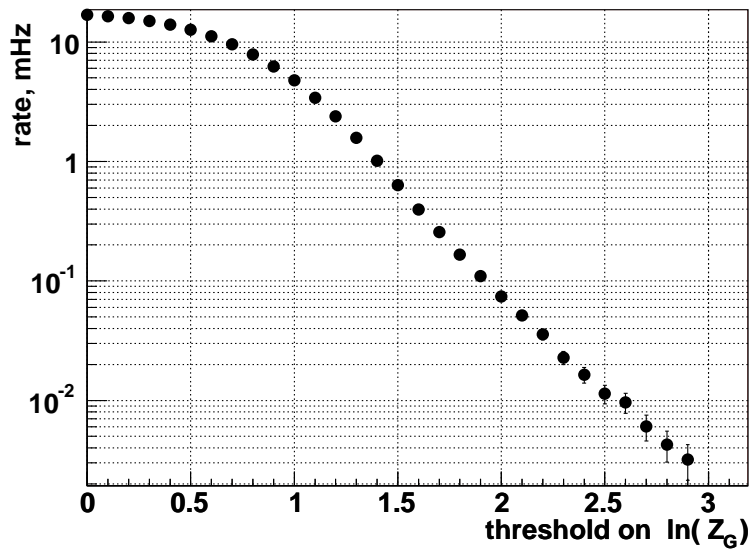}
  \end{center}
  \caption{The false alarm rate as a function of the threshold on the
  combined significance.}
  \label{fig:bgrate}
\end{figure}
\begin{figure}[ht]
   \centering\includegraphics[width=0.8\textwidth]{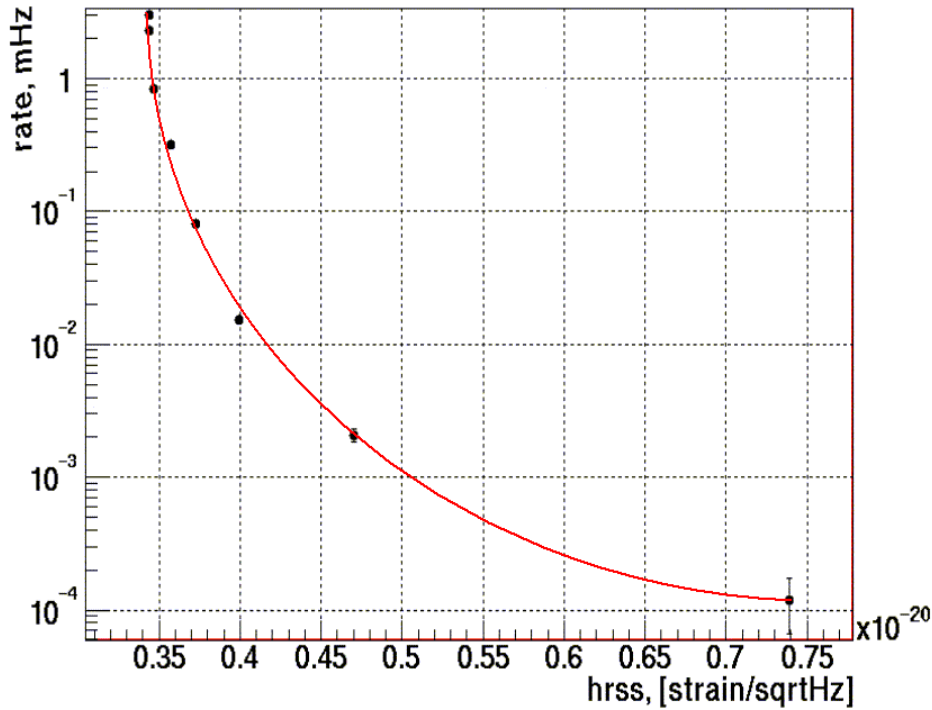}
   \caption{The dependence of the false alarm rate on the pipeline
   sensitivity for sine-Gaussian waveforms with $f=235$~Hz and
   $Q=9$. Solid circles correspond to the WaveBurst triggers with
   $\ln(Z_G)$ greater than 0, 0.5, 1.0, 1.25, 1.5, 1.7, 2.0, 2.5 
   (counting from the left). For convenience, the rate-hrss dependence 
   is approximated with a solid line.}
   \label{fig:roc}
\end{figure}

As shown in Figure~\ref{fig:bgrate}, the FA rate depends strongly on
the WaveBurst combined significance cut. 
Without compromising much the pipeline sensitivity,
the significance threshold is set to 5.47 (or $\ln(Z_G)>1.7$). 
At this threshold the FA rate is approximately $230~\mu{\mathrm{Hz}}$ 
for the WaveBurst triggers in the frequency band $64-4096$~Hz.
The rate is strongly dependent on the frequency band selected for the
analysis. For example, for the frequency
band below $1100$~Hz, the measured FA rate is approximately $15~\mu{\mathrm{Hz}}$.
Both the FA rate and the pipeline sensitivity depend on the significance
threshold. Varying the threshold, we can study the dependence of the FA
rate on the sensitivity, 
which is an important characteristic of the pipeline. Figure~\ref{fig:roc} 
shows the pipeline rate as a function of the sensitivity for  
frequency band below $1100$~Hz. 

\section{Simulation}
\label{s_sim}

For estimation of the WaveBurst detection efficiency we studied the
response of the analysis pipeline to simulated signals. Simulated
signals with different amplitudes were injected into the GW data
streams from three LIGO 
detectors. Then we applied the WaveBurst algorithm to find the injected
signals and thus estimated the pipeline sensitivity. 

Several ad-hoc and astrophysically motivated waveforms
were selected for injections. These included Gaussian, sine-Gaussian
waveforms and the simulated BH-BH merger waveforms described in
\cite{lazarus}. To make the injections as realistic as possible we
took into account 
the antenna pattern functions of the LIGO detectors and injected the
waveforms at random times, accounting for the delay in the arrival
time between the Hanford and Livingston sites.

\subsection{Simulation Procedure}

A general GW burst is comprised of two 
waveforms $h_{+}(t)$ and $h_{\times}(t)$ which represent two 
polarizations of the gravitational wave. The signal produced at the
output of the GW detector is a linear combination of
these waveforms:
\begin{equation}
   h(t) = F_{+} \, h_{+}(t) + F_{\times} \, h_{\times}(t) ,
\end{equation}
where $F_{+}$ and $F_{\times}$ are the antenna pattern functions. 
These functions depend on the source location in the sky 
(spherical angles $\theta$ and $\phi$) and its polarization angle $\psi$. 
To generate the antenna pattern functions for Hanford detectors, we explicitly 
construct the rotational transformation from the source coordinate 
frame to the Hanford frame. To obtain the $L1$ antenna pattern 
we apply a second transformation -- rotation from the Hanford to 
the Livingston detector frame. 

The amplitudes of injected signals are varied to obtain 
the sensitivity of the algorithm as a function of the injection strength.
For the Hanford data, the simulated signals are injected randomly in time at 
the average rate of 5 per minute. For the Livingston data, the same set of
waveforms is injected  
with time delays, uniquely defined by the source coordinates $\theta$
and $\phi$. The source coordinates are generated randomly, 
so the sources appear distributed uniformly on the sky ({\it all sky
simulation}) and their polarization
angles take random values between $0$ and $2\pi$. 
In addition, we injected waveforms with the same strength for both
sites, ignoring the antenna pattern functions. Although this
simulation does not correspond to any 
meaningful source population, it allows us to remove the contribution of the 
antenna pattern functions and estimate the best pipeline
sensitivity. Below we refer 
to this case as the {\it simulation with optimal orientation}.

\subsection{Pipeline sensitivity}

\begin{figure}[ht]
   \centering\includegraphics[width=0.8\textwidth]{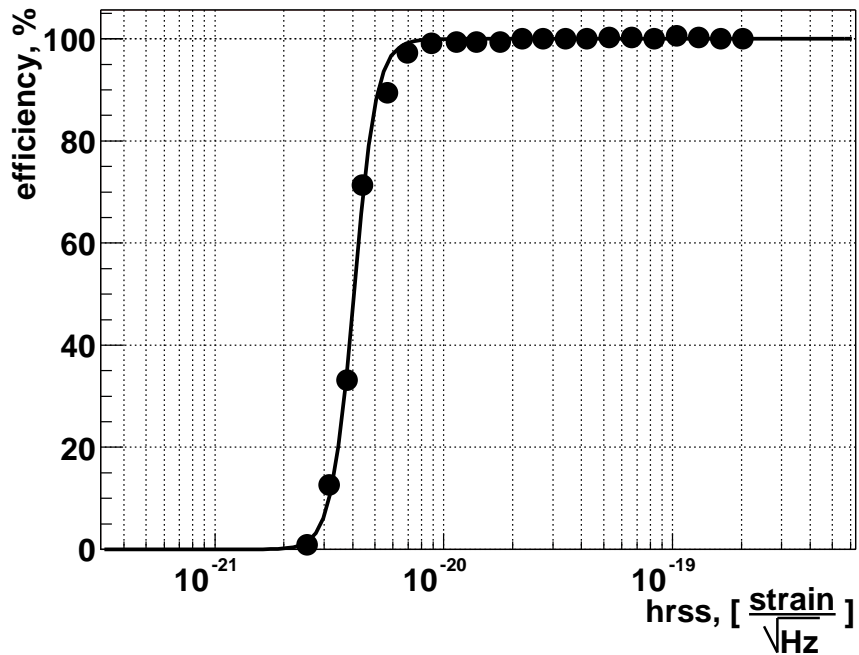}
   \caption{The detection efficiency for sine-Gaussian waveforms  
   ($f=235$~Hz, $Q=9$, optimal orientation).}
   \label{fig:sgeff}
\end{figure}

The detection efficiency is a function of the injected signal strength.
For a given signal strength, 
the detection efficiency is defined as the ratio of the number of 
detected waveforms to the total number of injected waveforms. 
We define the strength of an arbitrary burst signal as a root-sum-square 
strain amplitude \cite{LIGO_burst}:
\begin{equation}
   h_{\mathrm{rss}} = \left\{ \int \left[ h^2_{+}(t)+h^2_{\times}(t)
   \right]  dt \right\}^{1/2}. 
\end{equation}
For example, the detection efficiency curve as a function of 
$h_{\mathrm{rss}}$ 
is shown in Figure~\ref{fig:sgeff} for one of the simulated signals.
We determine the strength of signals detected with 50\% efficiency ($h_{50\%}$) 
and use it as a measure of the WaveBurst sensitivity. 
The results on the measured sensitivities are presented below for
different injected waveforms and the combined significance threshold of $1.7$.

The first class of injected waveforms corresponds to 
the GW bursts, which can be produced in the final plunge of two
coalescing black holes. 
Both $h_{+}(t)$ and $h_{\times}(t)$ 
waveforms have been obtained by the Lazarus group
\cite{lazarus,bhmerger}  as the result  
of numerical simulation of the 
merger and ringdown phases of a binary black hole system.
These waveforms are parametrized by the total mass of the binary system in
units of Solar mass $M_{\odot}$. The results on the pipeline
sensitivity are summarized in Table~\ref{table:bheff}.

\begin{table}[ht]
\begin{center}
   \begin{tabular}{r|rrrrrrrrrr}
   \hline\hline
   total mass ($M_{\odot}$) & 10 & 20 & 30 & 40 & 50 & 60 & 70 & 
       80 & 90 & 100 \\
   \hline
   sensitivity & 55 & 24 & 17 & 14 & 14 & 15 & 22 & 32 & 46 & 58\\
   \hline\hline
   \end{tabular}
   \caption{WaveBurst sensitivity (in units of 
   $10^{-21}$ $strain/\sqrt{\mathrm{Hz}}$) 
   for black hole waveforms (all sky simulation).}
\label{table:bheff}
\end{center}
\end{table}

Another class of injected waveforms is the sine-Gaussian wave \cite{LIGO_burst}
with the center frequency $f$ and the quality factor $Q = \sqrt{2}\pi f \tau$,
where $\tau$ defines the signal duration. 
The results of the pipeline sensitivity to the sine-Gaussian waveforms 
are summarized in  Table~\ref{table:sgeff}, assuming only one
polarization ($h_{+}$) of the injected signals. 
The sensitivity is greatest to the sine-Gaussian waveform with 
$f=235$~Hz, which lies in the frequency band
with maximum detector sensitivity. Note that the pipeline
has approximately the same sensitivity for two different types of 
sine-Gaussian waveforms: $Q=3$ and $Q=9$.

\begin{table}[ht]
   \begin{center}
   \begin{tabular}{c|rrrrrrrr} \hline\hline
freq. (Hz) &	100	&	153	& 235	&	361	&	554	&	849	&	1304	&	2000 \\ \hline
optimal & & & & & & & &  \\
$Q=3$	& 	28.2	&	11.5	&	4.8	&	5.5	&	8.2	&	14.0	&	25.4	&	52.2 \\
$Q=9$	&	27.3	&	18.8	&	4.0	&	5.1	&	6.7	&	11.5	&	20.3	&	51.2\\	\hline
all sky & & & & & & & & \\
$Q=3$	&	85.3	&	34.2	&	15.7	&	19.2	&	25.0	&	40.4	&	70.4	&	151.0 \\
$Q=9$	&	76.5	&	51.3	&	13.9	&	14.4   &	21.0	&	35.2	&	61.8	&	149.0\\ 
   \hline\hline
   \end{tabular}
   \caption{WaveBurst sensitivity (in units of $10^{-21}$
   $strain/\sqrt{\mathrm{Hz}}$)   
   for sine-Gaussian waveforms (optimal orientation and all sky simulations).}
   \label{table:sgeff}
\end{center}
\end{table}

Finally, we estimated the WaveBurst sensitivity to the pure Gaussian
waveforms \cite{LIGO_burst}, 
which are characterized by the duration $\tau$ only. For these signals one 
polarization ($h_{+}$) and optimal orientation are assumed.
The results are summarized in Table~\ref{table:gaeff}. 

\begin{table}[ht]
\begin{center}
\begin{tabular}{c|r|r|r|r|r}
\hline\hline
$\tau$ (ms)   & 0.1  & 0.5	&1.0	&2.5	&4.0	\\
sensitivity  & 14.4 & 8.2	&9.4	&49.2	&154.0	\\
\hline\hline
\end{tabular}
\caption{WaveBurst sensitivity (in units of 
$10^{-21}$ $strain/\sqrt{\mathrm{Hz}}$) for
Gaussian waveforms (optimal orientation).}
\label{table:gaeff}
\end{center}
\end{table}

\section{Conclusion}

WaveBurst is a novel method for detection of gravitational wave
bursts. It works in  the wavelet domain and allows detection of a
wide class of GW bursts by using  a large bank of wavelet packets.
Using the S2 LIGO playground data we evaluated the performance of the
WaveBurst  data analysis pipeline.
The pipeline sensitivity is limited by thresholds on: the wavelet
amplitudes (defined by the ETG parameters $P$ and $\eta$), 
the cluster excess power ratio, and the combined
significance of the triple coincidence events.  The maximum pipeline
sensitivity is approximately $4 \times 10^{-21}
strain/\sqrt{\mathrm{Hz}}$ for sine-Gaussian  signals
corresponding to optimal orientation of sources with respect to
the LIGO detectors.  Averaged over the entire sky, the pipeline strain
sensitivity becomes  $14 \times 10^{-21}
strain/\sqrt{\mathrm{Hz}}$. The false alarm rate of the pipeline is
dominated by accidental triple coincidence events  produced by
fluctuations in the detector noise. Using time shift analysis we
estimated the WaveBurst false alarm rate as a function of the combined
significance threshold. For the threshold of $1.7$,
the false alarm rates are $15~\mu{\mathrm{Hz}}$ and
$230~\mu{\mathrm{Hz}}$ for frequency bands below $1100$~Hz and
$4096$~Hz respectively. Thus we have shown that the WaveBurst algorithm 
has low false alarm rates and high sensitivity to simulated burst waveforms.

\ack

We thank Peter Saulson for comments on the paper.  We also thank the
LIGO laboratory and the LIGO Scientific Collaboration  which enabled
this work by constructing and operating the LIGO interferometers.  We
gratefully acknowledge the support of the United States National
Science Foundation for the construction and operation of the LIGO
Laboratory and the Particle Physics and Astronomy Research Council of
the United Kingdom, the Max-Planck-Society and the State of
Niedersachsen/Germany for support of the construction and operation of
the GEO600 detector. We also  gratefully acknowledge the support of
the research by these agencies and by the Australian Research Council,
the Natural Sciences and Engineering Research  Council of Canada, the
Council of Scientific and Industrial Research of India, the Department
of Science and Technology of India, the Spanish Ministerio de Ciencia
y Tecnologia, the John Simon Guggenheim Foundation, the David and
Lucile Packard Foundation, the Research Corporation, and the Alfred
P. Sloan Foundation. This work is also supported by the US National
Science Foundation grants PHY-0244902 and PHY-0070854. The paper has
been assigned the LIGO Laboratory document number LIGO-P040012-00-Z.


\section*{References}

\begin{thebibliography}{00}
%
\bibitem{LIGO} Abbott B \etal 2004 \NIM {\bf 517} 154
%
\bibitem{GEO} Willke B \etal 2002 \CQG {\bf 19} 1377
%
\bibitem{VIRGO} Acernese F \etal 2002 \CQG {\bf 19} 1421
%
\bibitem{TAMA} Tagoshi H \etal 2001 \PR D {\bf 63} 062001
%
\bibitem{LIGO_burst} Abbott B \etal 2004 \PR D {\bf 69} 102001
%
\bibitem{tf1} Anderson~W and Balasubramanian~R 1999 \PR D {\bf 60} 102001
%
\bibitem{tf2} Mohanty~S 2000 \PR D {\bf 61} 122002
%
\bibitem{power} Anderson~W, Brady~P, Creighton~J and Flanagan~E
                 2001 \PR D {\bf 63} 042003
%
\bibitem{tf3} Sylvestre J 2002 \PR D {\bf 66} 102004
%
\bibitem{method} Klimenko S and Mitselmakher G 2004 {\it A wavelet 
method for detection of gravitational wave bursts} (in this volume)
%
\bibitem{vidakovic} Vidakovic B 1999 {\it Statistical modeling by
wavelets} (New York: Wiley Interscience)
%
\bibitem{packets1} Coifman R, Meyer Y and Wickerhauser M 
1992 {\it Wavelet analysis and signal processing} In Ruskai M \etal
{\it Wavelets and their applicaltions} pages 153-178 (Jones and 
Bartlett Publishers, Sudbury, MA)
%
\bibitem{ingrid} Daubechies I 1992 {\it Ten lectures on wavelets} 
(Philadelphia: SIAM)
%
\bibitem{lazarus} Baker J, Companelli M and Lousto C 
2002 \PR  D {\bf 65} 044001
%
\bibitem{bhmerger} Baker J, Companelli M, Lousto C and Takahashi R 
2002 \PR  D {\bf 65} 124012


\end{thebibliography}
\end{document}